\documentclass[aps,prb,reprint,superscriptaddress,twocolumn,floatfix]{revtex4-2}
\usepackage[T1]{fontenc}
\usepackage{textcomp}
\usepackage{amsmath}
\usepackage{amsfonts}
\usepackage{amssymb}
\usepackage{graphicx}
\usepackage{siunitx}
\usepackage{setspace}
\usepackage{xcolor}
\usepackage[colorlinks=true,linkcolor=black,citecolor=blue,urlcolor=black]{hyperref}

\newcommand{\revision}[1]{\textcolor{black}{#1}}

\begin{document}

\title{\revision{Multimode} Single-Ring Photonic Molecule}
\author{Jinsheng Lu}
\email{jlu@seas.harvard.edu}
\affiliation{Harvard John A. Paulson School of Engineering and Applied Sciences, 9 Oxford Street, Cambridge, MA 02138, USA}

\author{Ileana-Cristina Benea-Chelmus}
\affiliation{Harvard John A. Paulson School of Engineering and Applied Sciences, 9 Oxford Street, Cambridge, MA 02138, USA}
\affiliation{Hybrid Photonics Laboratory, École Polytechnique Fédérale de Lausanne (EPFL),CH-1015, Switzerland}

\author{Vincent Ginis}
\affiliation{Harvard John A. Paulson School of Engineering and Applied Sciences, 9 Oxford Street, Cambridge, MA 02138, USA}
\affiliation{Data Lab / Applied Physics, Vrije Universiteit Brussel 1050 Brussel, Belgium}

\author{Marcus Ossiander}
\affiliation{Harvard John A. Paulson School of Engineering and Applied Sciences, 9 Oxford Street, Cambridge, MA 02138, USA}
\affiliation{Institute of Experimental Physics, Graz University of Technology, 8010 Graz, Austria}

\author{Danilo Shchepanovich}
\affiliation{Harvard John A. Paulson School of Engineering and Applied Sciences, 9 Oxford Street, Cambridge, MA 02138, USA}

\author{Federico Capasso}
\email{capasso@seas.harvard.edu}
\affiliation{Harvard John A. Paulson School of Engineering and Applied Sciences, 9 Oxford Street, Cambridge, MA 02138, USA}
\begin{abstract}
Photonic molecules can mimic interactions of atomic energy levels, offering new ways to manipulate cavity eigenstates. Current methods using evanescent coupling of multiple cavities face challenges in scalability, flexibility, and coupling control, especially for complex systems. Here we introduce a new method that uses a single multimode optical ring resonator to create photonic molecules. Our design uses multiple waveguide transverse modes in one resonator, providing flexibility to engineer complex interactions without typical coupling constraints. We demonstrate arbitrary inter-mode coupling through transmissive mode converters, allowing precise tuning of resonance splitting and intrinsic losses. This approach enables selective bright-dark mode pair generation and the exploration of novel photonic phenomena such as exceptional points. This multimode photonic molecule overcomes traditional limitations and offers new possibilities for integrated photonic circuits, optical processing, and studies in non-Hermitian and nonlinear photonics.
\end{abstract}

\maketitle


Coupled optical resonators with engineered photon interactions, termed photonic molecules \cite{Bayer1998optical, Rakovich2010photonic, Haddadi2014photonic, Chremmos2010photonic, Liao2020photonic}, have attracted considerable interest because of their unique ability to emulate interactions of atomic energy levels \cite{zhang2019electronically}. This analogy has provided a robust platform for exploring fundamental phenomena such as quantum optics \cite{PhysRevLett.104.183601, PhysRevA.95.013815, PhysRevLett.98.213904, PhysRevLett.102.173902,bose2014all}, coherent energy transfer \cite{sato2012strong,bose2014all}, parity-time symmetry \cite{hamel2015spontaneous,hodaei2014parity,chang2014parity,miri2019exceptional,wang2023non}, topological photonic systems \cite{sridhar2024quantized,dutt2020higher}, and nonlinear interactions \cite{rebolledo2023platicon,huHighefficiencyBroadbandOnchip2022,li2025broadband,tikanProtectedGenerationDissipative2022}. These systems enable precise control over the photon energy and phase, which is essential for applications in advanced optical modulation and switching \cite{bose2014all, zhang2019electronically, ZhaoOE2015, NozakiOE2013, zhang2015ultra}, quantum information processing \cite{Dousse2010ultrabright, Zhang2021squeezed, PhysRevLett.92.083901}, and sensing  \cite{hodaei2017enhanced,xu2023breaking,ta2014coupled,cardador2018photonic, ZhangOE2011, Xu2024single}.

\begin{figure}[t]
\includegraphics[width = 0.45\textwidth]{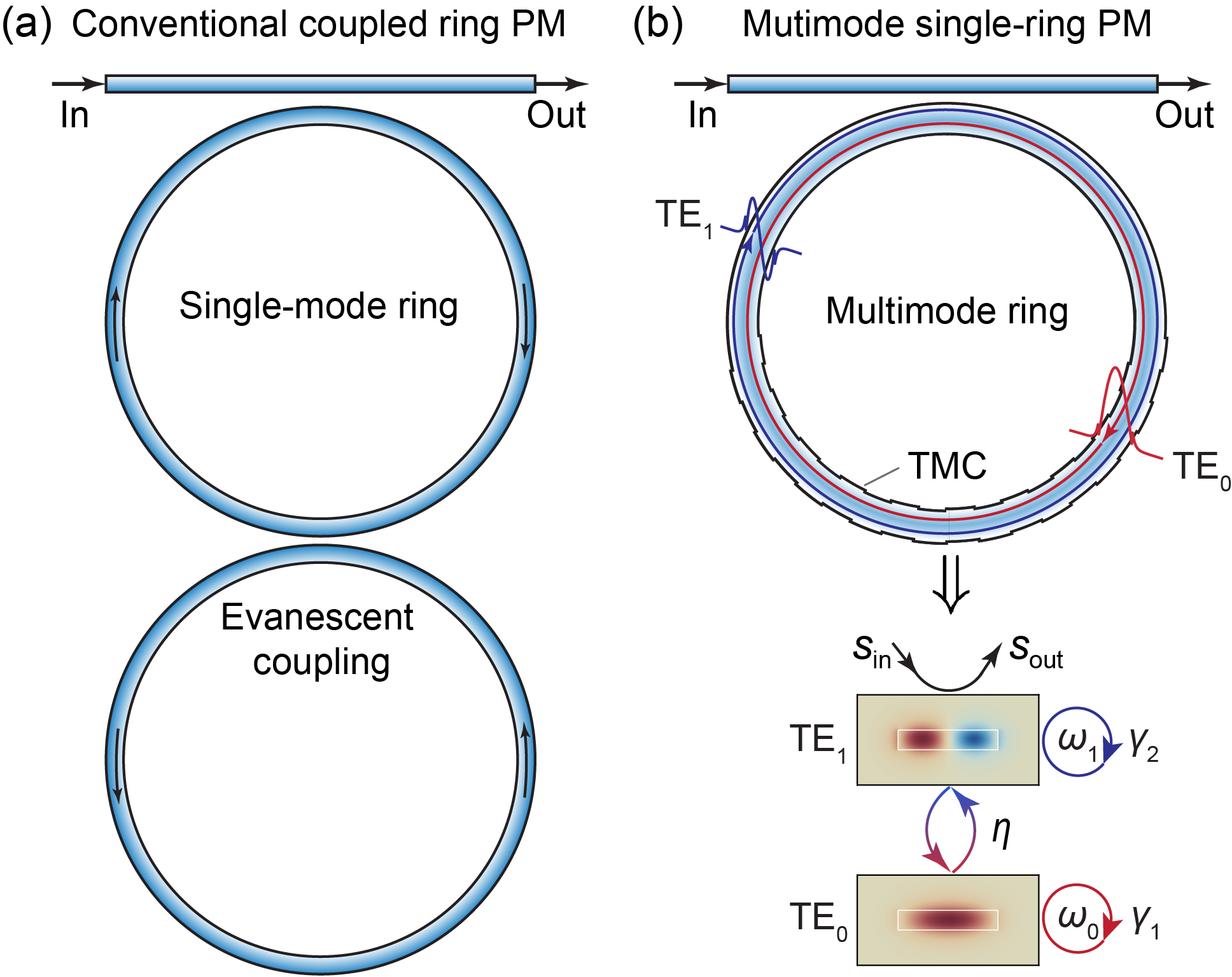}
\caption{\label{fig:fig0} \revision{Conventional and multimode photonic molecules. (a) A conventional photonic molecule (PM) formed by two evanescently coupled rings.} (b) A multimode single-ring photonic molecule that exploits multiple waveguide transverse modes and transmissive mode converters (TMCs). The schematic illustrates the coupling dynamics between the $\mathrm{TE_0}$ and $\mathrm{TE_1}$ modes within the multimode architecture.}
\end{figure}

Traditionally, photonic molecules are implemented by coupled microresonators  \cite{Rakovich2010photonic,zhang2019electronically,siegle2017photonic,peng2012photonic} or photonic crystal cavities \cite{zhu2025full,ji2023thickness}, based on evanescent coupling (Fig. \ref{fig:fig0}(a)). This near-field interaction inherently limits coupling to adjacent resonators, making arbitrary or long-range coupling challenging \cite{absil2001vertically,little1997microring}. \revision{Single-resonator implementations based on Bragg gratings that couple clockwise and counterclockwise modes have also been explored \cite{Goede2021modesplitting}. However, because these approaches primarily exploit coupling within the same transverse mode, they support only a two-level energy structure.} Therefore, these configurations face limitations in scalability, flexible mode interaction, and precise coupling strength control \cite{siegle2017photonic,tao2024versatile,smith2020active}. The implementation and control of more complex energy-level structures \cite{PhysRevA.95.033847,wang2023integrated,van2007synthesis} is challenging because it requires the coupling of multiple resonators in carefully designed configurations. 

\begin{figure}[t]
\includegraphics[width = 0.45\textwidth]{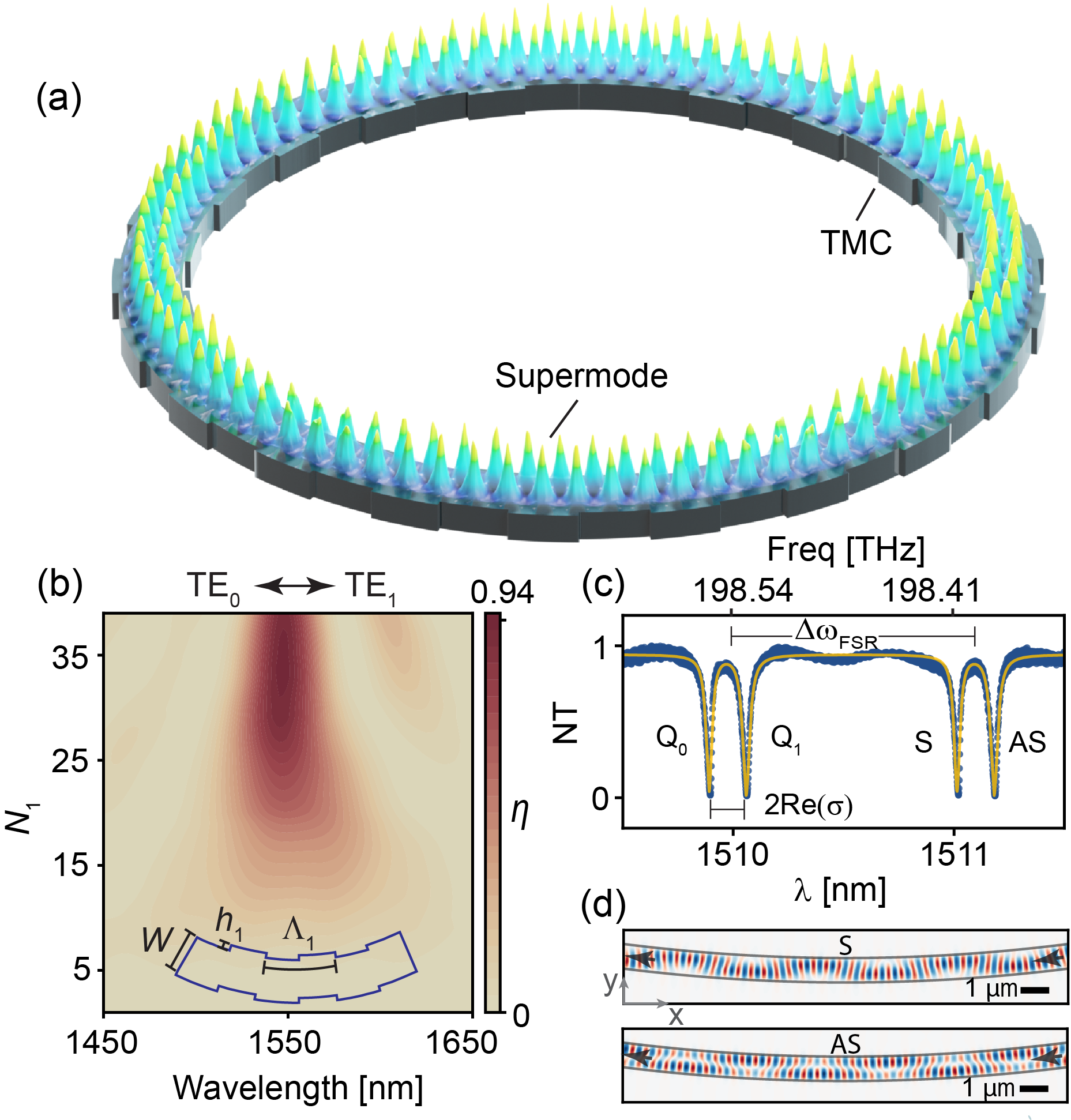}
\caption{\label{fig:fig1} Properties of a multimode single-ring photonic molecule.
(a) Artistic rendering of a corrugated multimode ring resonator, overlaid with a simulated supermode intensity profile (bus waveguide not shown). (b) Simulated power conversion efficiency $\eta$ versus wavelength and the grating period number $N_1$ of the $\mathrm{TE_0}$-$\mathrm{TE_1}$ TMC. Inset: schematic of an asymmetric-grating TMC segment. Device parameters: Ring waveguide width $W$ = 1100 nm, radius 80 $\mu$m, thickness 220 nm, corrugation depth $h_1$ = 20 nm, period $\Lambda$ = 6300 nm. (c) Measured normalized transmission (NT) spectrum of a multimode single-ring photonic molecule (yellow curve: fit using Eq. 2). The free spectral range is $\Delta \omega_\text{FSR} = 2\pi\times147.7$ GHz. Two hybridized modes are split by $2\text{Re}(\sigma) = 2\pi\times21.8$ GHz, with linewidths $\gamma_0 = 2\pi\times1.5$ GHz and $\gamma_1 = 2\pi\times2.3$ GHz, corresponding to loaded optical quality factors of $Q_0 = 6.5\times10^4$ and $Q_1 = 4.3\times10^4$. (d) Symmetric (S) and antisymmetric (AS) optical modes profiles, represented by $\mathrm{Re(H_z)}$. Arrows indicate propagation direction. }
\end{figure}

In this work, we propose and experimentally demonstrate a novel class of photonic molecules based on a single multimode optical ring resonator (Fig. \ref{fig:fig0}(b)). By leveraging multiple waveguide transverse modes supported within a single cavity, our compact design achieves functions similar to traditional multi-resonator systems. We develop a generalized analytic framework for this system with an arbitrary number of modes and mode couplings using the transfer function method. Our multimode approach enables arbitrary and precise control of the inter-mode coupling via transmissive mode converters \cite{doi:10.1126/sciadv.adt4154} without geometric restrictions. Moreover, different transverse modes exhibit distinct mode profiles, effective indices (dispersion), and intrinsic losses, allowing us to independently engineer the real and imaginary parts of the system's eigenvalues. This capability facilitates the exploration of richer mode dynamics, including bright-dark mode pairs \cite{ji2024multimodality}, exotic phenomena such as exceptional points \cite{chen2017exceptional,miri2019exceptional,brandstetter2014reversing,caselliGeneralizedFanoLineshapes2018}, and advanced dispersion engineering, particularly on nonlinear integrated platforms such as lithium niobate or silicon nitride  \cite{helgason2021dissipative,yuan2023soliton,xueSuperefficientTemporalSolitons2019,zangLaserPowerConsumption2025,liThermalTuningMode2022,helgason2023surpassing}.

We design a simple single-ring photonic molecule based on a multimode silicon ring resonator, as illustrated in Fig. \ref{fig:fig0}(b) and Fig. \ref{fig:fig1}(a). A transmissive mode converter (TMC) integrated in the ring uses periodic corrugations characterized by a long grating period ($\Lambda_1$) and shallow corrugation depth. The TMC facilitates efficient co-directional coupling between the $\mathrm{TE}_0$ and $\mathrm{TE}_1$ modes (Sec. S1). The grating period $\Lambda_1$ is selected to satisfy the phase matching condition 
$
\Lambda_1 = \frac{\lambda_0}{\Delta n_{\text{eff},01}}
$,
where $\lambda_0$ represents the vacuum wavelength and $\Delta n_{\text{eff},01}$ is the effective refractive index difference between the $\mathrm{TE}_0$ and $\mathrm{TE}_1$ modes. The power conversion efficiency $\eta$ of the TMC is determined by \cite{doi:10.1126/sciadv.adt4154}
$
\eta = \frac{\kappa^2}{\kappa^2+\delta^2}\sin^2{(\sqrt{\kappa^2+\delta^2}N_1\Lambda_1)}
$,
where $\kappa$ is the coupling coefficient, $\delta$ is the phase mismatch between the two transverse modes, and $N_1$ is the grating period number. Simulation results indicate that the power conversion efficiency between the $\mathrm{TE}_0$ and $\mathrm{TE}_1$ modes increases with the number of grating periods ($N_1$), reaching a maximum efficiency of 0.94 at $N_1 = 33$, as depicted in Fig. \ref{fig:fig1}(b). \revision{The reflection from the gratings is negligible (less than 0.002 when $N_1 = 33$).} These simulation results align closely with theoretical predictions (Fig. S1). The maximum efficiency is limited by the bending loss. 

We use a bus waveguide with a width of 505 nm and a gap of 200 nm to selectively fulfill the phase matching condition between the fundamental mode of the bus waveguide and the $\mathrm{TE}_1$ mode in the ring waveguide (Fig. \ref{fig:fig0}(b)). This selective phase matching enables coupling the $\mathrm{TE}_1$ mode into and out of the ring (i.e., $\mathrm{TE}_1$ is a bright mode), while the other modes in the ring become dark modes (that cannot be coupled out). Fig. \ref{fig:fig0}(b) illustrates the mode dynamics in the single-ring photonic molecule. It supports a pair of well-defined optical energy levels, as evidenced by the normalized transmission spectrum shown in Fig. \ref{fig:fig1}(c), resulting from mode coupling in the TMC. These two energy levels correspond to symmetric (S) and antisymmetric (AS) optical modes (Fig. \ref{fig:fig1}(d), Fig. S2, Fig. S25), generated because the two overlapping transverse modes are in- or out-of-phase. The simulated field intensity distribution in the ring, artistically rendered in Fig. \ref{fig:fig1}(a), illustrates a resonant supermode. We refer to this resonance as a supermode because a new roundtrip mode is formed that simultaneously includes both transverse modes (Fig. S3, Fig. S24, and Fig. S26). We will explain the supermode in more detail later.

\begin{figure}[t]
\includegraphics[width = 0.5\textwidth]{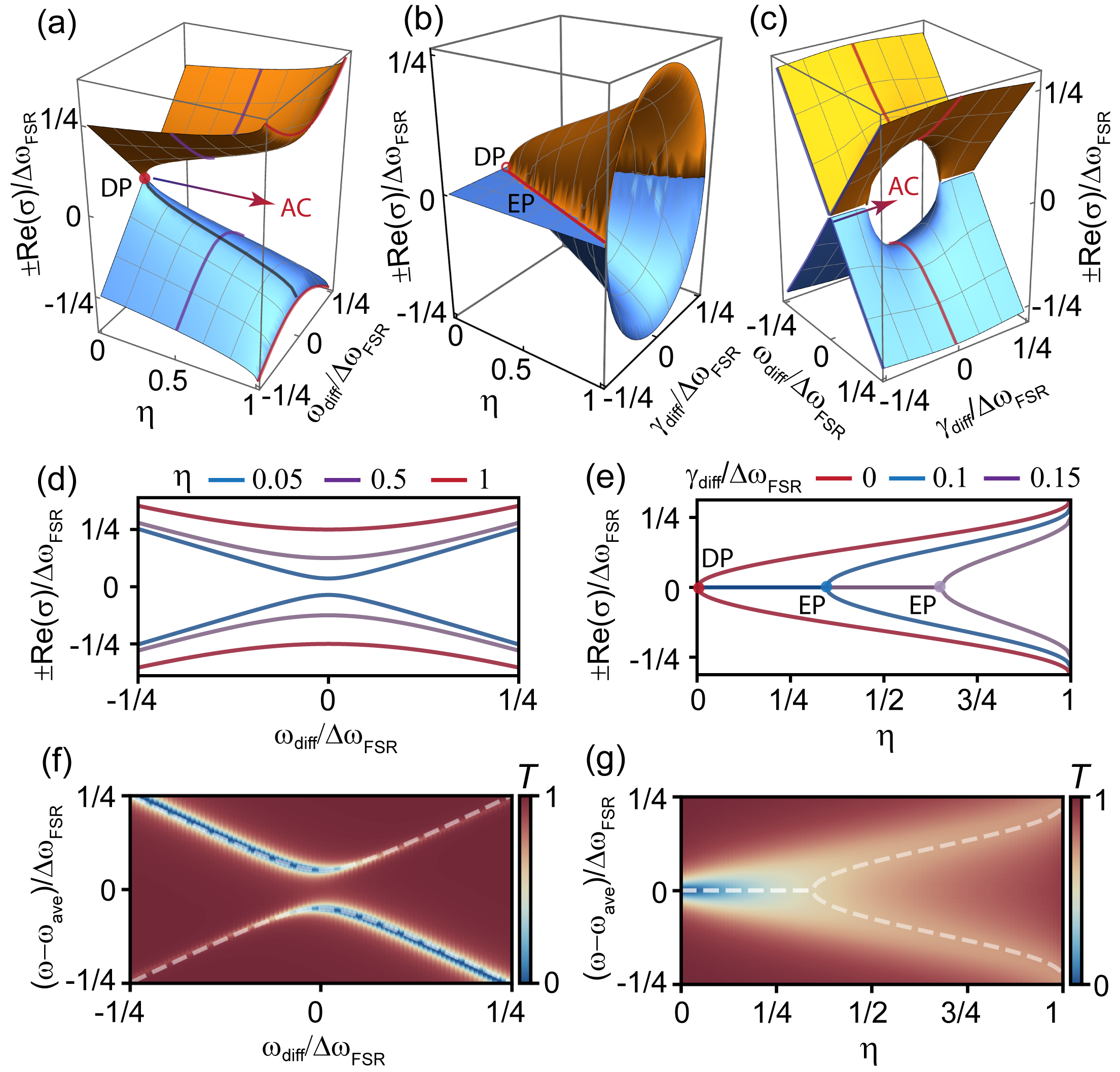}
\centering
\caption{\label{fig:fig2} Theoretical analysis of resonance frequency splitting behavior in a multimode single-ring photonic molecule.
Calculated real part of the eigenvalue splitting (normalized to the free spectral range), $\pm\mathrm{Re}(\sigma)/ \Delta\omega_\text{FSR}$, as a function of various system parameters using the coupled-mode theory. 
(a) $\gamma_\text{diff}$ = 0. (b) $\omega_\text{diff} $ = 0. (c) $\eta$ = 0.5. (d) $\gamma_\text{diff}$ = 0 and $\eta$ = 0.05 (blue curve), 0.5 (purple curve), and 1 (red curve). (e) $\omega_\text{diff} $ = 0 and $\gamma_\text{diff}/\Delta\omega_\text{FSR}$ = 0 (red curve), 0.1 (blue curve), and 0.15 (purple curve). DP: diabolic point, AC: anti-crossing, EP: exceptional point. (f, g) Analytically calculated transmission spectra of the single-ring photonic molecule system obtained using the transfer matrix method. (f) $\eta$ = 0.1 and $\gamma_\text{diff}$ = 0. (g)  $\gamma_\text{diff}/\Delta\omega_\text{FSR}$ = 0.1 and $\omega_\text{diff} $ = 0. The white dashed lines indicate fits derived from coupled-mode theory.}
\end{figure}

\begin{figure}[t]
\includegraphics[width = 0.5\textwidth]{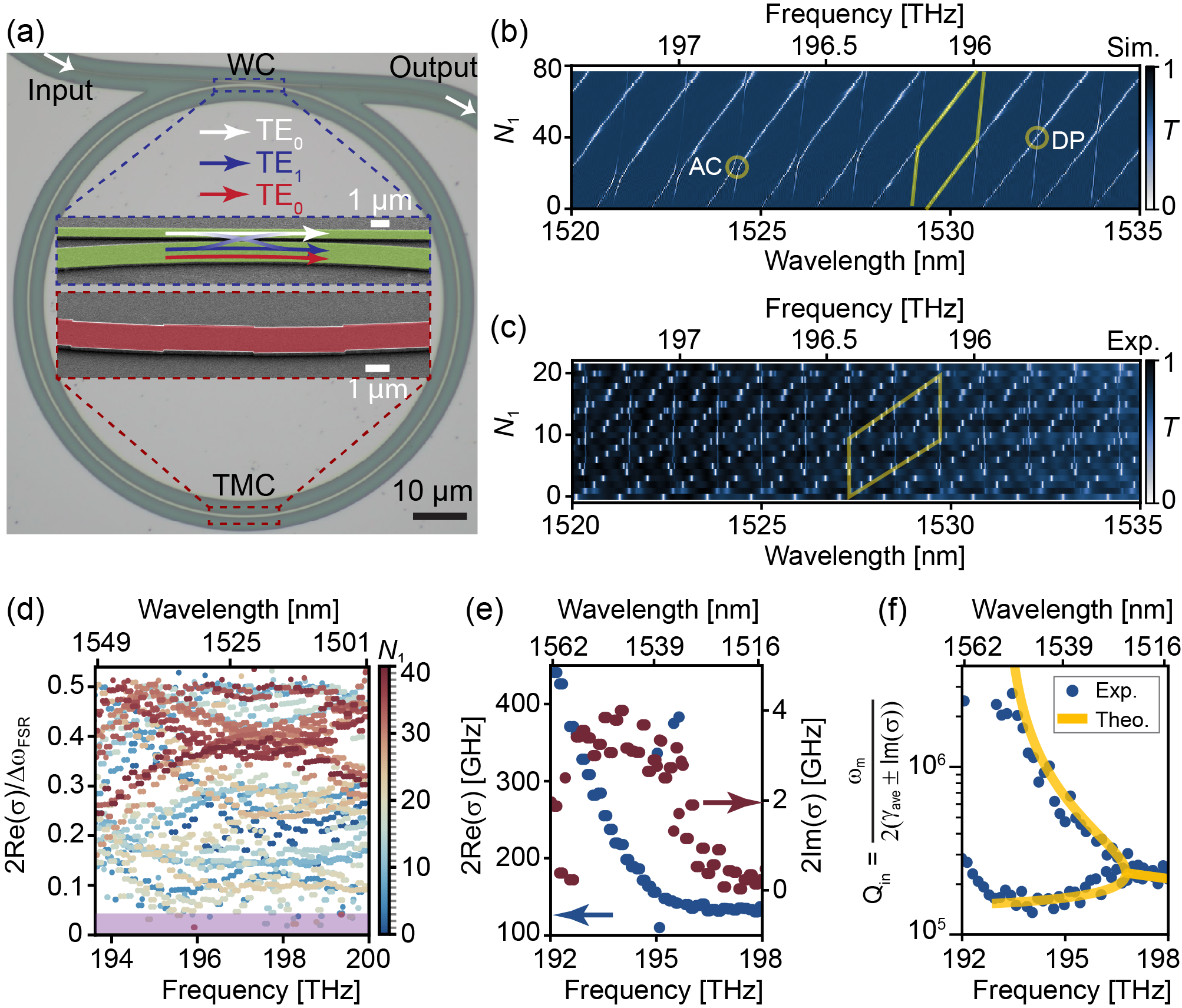}
\caption{\label{fig:fig3} Experimental demonstration of a multimode single-ring photonic molecule. (a) Optical microscope image of a fabricated single-ring multimode resonator. Insets show tilted scanning electron microscope images for the waveguide coupler (WC) and TMC. The device parameters are the same as in Fig. 2. (b,c) Simulated and experimentally measured normalized transmission spectrum at the bus waveguide output of a single-ring photonic molecule when varying $N_1$. The yellow circles highlight the positions of anti-crossing (AC) and diabolic point (DP). (d) Experimentally measured resonance frequency splitting normalized to the free spectral range ($\mathrm{2Re(\sigma)/\Delta\omega_\text{FSR}}$) as a function of the resonance frequency when varying $N_1$. (e,f) Measured real (blue dots) and imaginary (red dots) components of the eigenvalue splitting (e) and the corresponding intrinsic quality factors (f) of the single-ring photonic molecule system when $N_1$ = 9, as a function of the resonance frequency.}
\end{figure}

We explore the two-level system of this single-ring photonic molecule using coupled-mode theory (Sec. S2 and S3). The eigenvalues of this system are \cite{miri2019exceptional}
$
\text{E} = \omega_\text{ave} - \mathrm{i}\gamma_\text{ave} \pm \sigma
$, 
where $\omega_\text{ave} = \frac{\omega_{0} + \omega_{1}}{2}$ and $\gamma_\text{ave} = \frac{\gamma_0+\gamma_1}{2}$ represent the average of the resonance frequencies and the decay rates of the amplitudes of the two transverse modes. The eigenvalue splitting $\sigma$ can be expressed as
\begin{equation} \label{eq1}
   \sigma = \Delta\omega_\text{FSR} \sqrt{\left(\frac{\arcsin{\sqrt{\eta}}}{2\pi}\right)^2+\left(\frac{\omega_\text{diff}}{\Delta\omega_\text{FSR}}+\mathrm{i}\frac{\gamma_\text{diff}}{\Delta\omega_\text{FSR}}\right)^2} 
\end{equation}
where $\omega_\text{diff} = \frac{1}{2}(\omega_{1} - \omega_{0})$ and $\gamma_\text{diff} = \frac{1}{2}(\gamma_1-\gamma_0)$ denote the differences of the resonance frequencies and the decay rates of the two waveguide transverse modes. The effective free spectral range (FSR) of this two-mode system is defined as $\Delta\omega_\text{FSR} = \sqrt{\Delta\omega_\text{FSR,1}\Delta\omega_\text{FSR,2}}$, with the modes' individual FSRs $\Delta\omega_{\text{FSR},m} = \frac{2\pi c}{n_{\text{g},m}L}$. Here, $n_{\text{g},m}$ is the group index of the transverse mode $m$ ($m = 0, 1$), $c$ is the speed of light in vacuum, and $L$ is the length of the resonator.

Eq. \ref{eq1} describes the real (resonance frequency) and imaginary (decay rate) eigenvalue splitting components. Fig. \ref{fig:fig2}(a-e) analyzes the evolution of the eigenvalues' real components within a parameter space defined by the power conversion efficiency $\eta$, resonance frequency difference $\omega_\text{diff}$, and decay rate difference $\gamma_\text{diff}$, normalized to the FSR $\Delta\omega_\text{FSR}$. The corresponding imaginary parts are shown in Figs. S4 and S5. A typical resonance frequency splitting as a function of $\eta$ occurs at $\gamma_\text{diff} = 0$ and $\omega_\text{diff} = 0$, creating a diabolic point (DP) as seen in Fig. \ref{fig:fig2}(a,e).  Increasing the power conversion efficiency $\eta$ results in anti-crossing (AC) energy levels when varying $\omega_\text{diff}$, as illustrated in Fig. \ref{fig:fig2}(a,c,d). Exceptional points occur when varying $\eta$ with non-zero $\gamma_\text{diff}$, as shown in Fig. \ref{fig:fig2}(b,e). This phenomenon arises naturally due to the different loss rates ($\gamma_\text{diff} \neq 0$) of the $\mathrm{TE}_0$ and $\mathrm{TE}_1$ modes. When full mode conversion is achieved ($\eta = 1$), the eigenvalue splitting satisfies $2\text{Re}(\sigma) = \frac{1}{2}\Delta \omega_\text{FSR}$, indicating that the FSR is halved (assuming $\omega_\text{diff} = 0$ and $\gamma_\text{diff} = 0$). \revision{The energy picture based on coupled-mode theory described above provides intuitive physical insights into the system's mode dynamics and accurately captures the behavior in the weak and strong coupling regimes (Sec. S3 and Fig. S7)}.

Next, we employ the transfer function method to perform a rigorous and broadband analysis of the multimode single-ring photonic molecule system. Using this method, we developed a generalized analytical framework capable of handling single-ring photonic systems with an arbitrary number of transverse modes and mode coupling interactions (Sec. S4). For the simplified scenario involving only two transverse modes, the normalized output field amplitude can be expressed as:
\begin{equation}\label{eq2}
    \frac{s_\text{out}}{s_\text{in}} = -r + \frac{t^2\phi_1(\tau -\phi_0)}{1+r\phi_0\phi_1-r\tau\phi_1-\tau\phi_0}
\end{equation}
where $t$ (with $|r|^2+|t|^2 = 1$) indicates the amplitude coupling efficiency from the bus waveguide. The power conversion efficiency ($\eta$) of the $\mathrm{TE}_0$-$\mathrm{TE}_1$ TMC satisfies $\tau^2 = 1 - \eta$. The roundtrip propagation factor for transverse mode $m$ is defined as $\phi_m= \alpha_{\text{rt},m}e^{\mathrm{i}\frac{\omega}{c}n_{\text{eff},m}L}$, incorporating both the roundtrip phase term $e^{\mathrm{i}\frac{\omega}{c}n_{\text{eff},m}L}$ and the roundtrip field attenuation factor $\alpha_{\text{rt},m}$. In this expression, $n_{\text{eff},m}$ is the effective index, and $\omega$ denotes the angular frequency. The intrinsic loss rates of the transverse modes $\gamma_{\text{in},m} = -\ln{\alpha_{\text{rt},m}}/T_{\text{rt},m}$ are determined by  the round-trip group delays $T_{\text{rt},m} = {n_{\text{g},m}L}/{c}$, where $n_{\text{g},m}$ is the group index. The mode indices are specified such that $m = 0$ corresponds to $\mathrm{TE}_0$, and $m = 1$ corresponds to $\mathrm{TE}_1$. The normalized transmission spectrum is obtained by $T = |\frac{s_\text{out}}{s_\text{in}}|^2$. Results calculated using Eq. \ref{eq2}, as illustrated in Fig. \ref{fig:fig2}(f,g) and Figs. S6 and S7 for varying $\omega_\text{diff}$ and $\eta$, closely match the predictions provided by coupled-mode theory. Equation 2 also effectively fits the experimentally measured transmission spectra and accurately captures the resonances' linewidths (Fig. \ref{fig:fig1}(c) and Figs. S13–S15).

\begin{figure}[t]
\includegraphics[width = 0.5\textwidth]{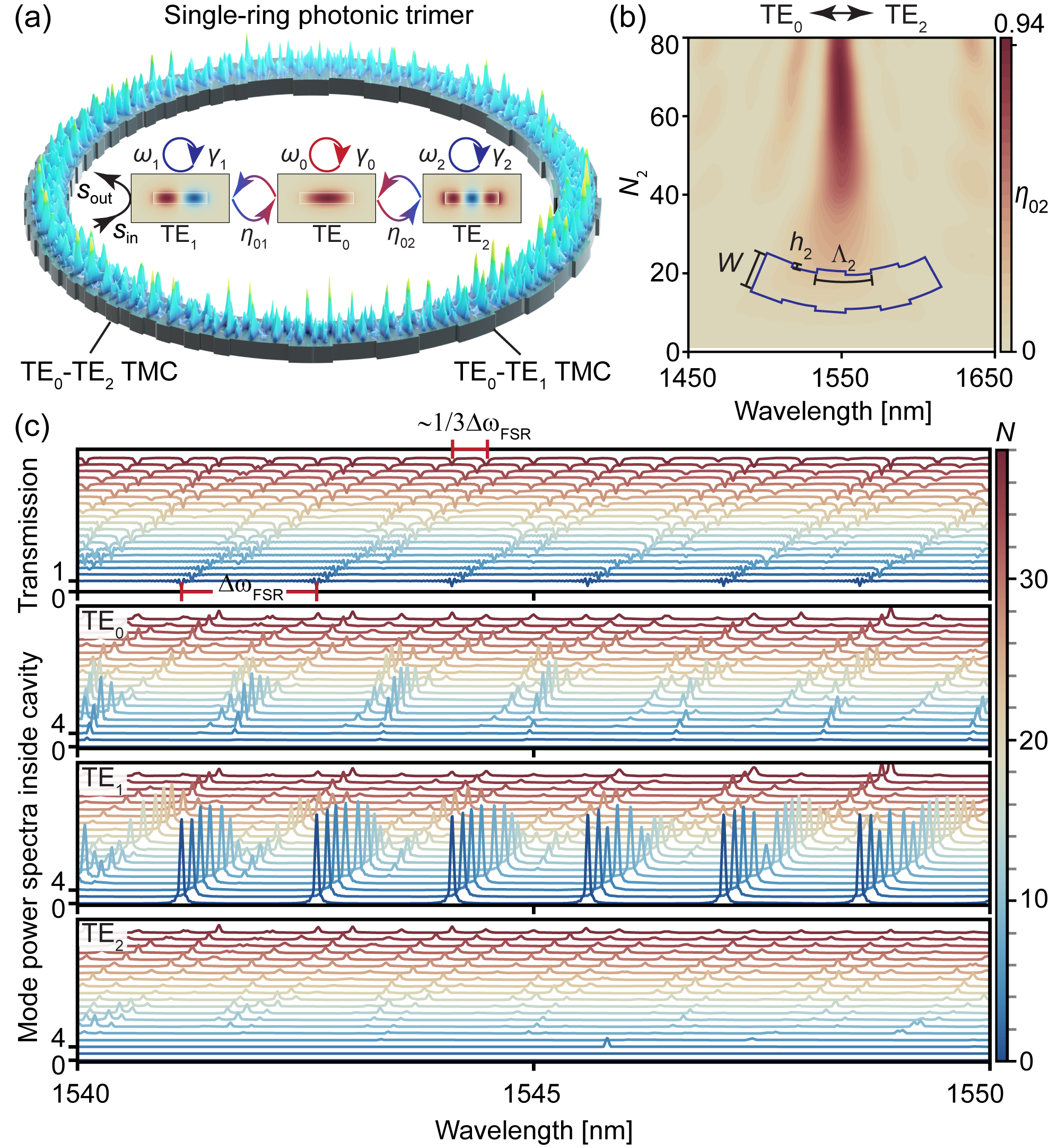}
\caption{\label{fig:fig4} Single-ring photonic trimer. (a) Schematic illustration of a single-ring photonic trimer and its mode coupling dynamics (insert), consisting of a ring resonator integrated with two TMCs for $\mathrm{TE_0}$-$\mathrm{TE_1}$ and $\mathrm{TE_0}$-$\mathrm{TE_2}$ mode interactions. (b) Simulated power conversion efficiency $\eta_{02}$ as a function of operating wavelength and the grating period number $N_2$ of the $\mathrm{TE_0}$-$\mathrm{TE_2}$ TMC. Inset: schematic of a segment of the $\mathrm{TE}_0$–$\mathrm{TE}_2$ TMC featuring symmetric gratings ($h_2$ = 15 nm, $\Lambda$ = 2250 nm). (c) Simulated transmission spectrum at the bus waveguide output, along with the intracavity power spectra of $\mathrm{TE}_0$, $\mathrm{TE}_1$, and $\mathrm{TE}_2$ modes when varying $N$, with the grating period numbers of the $\mathrm{TE_0}$–$\mathrm{TE_1}$ TMC and $\mathrm{TE_0}$–$\mathrm{TE_2}$ TMC set as $N_1 = N$ and $N_2 = 2N$, respectively.}
\end{figure}

Importantly, when complete mode conversion occurs ($\eta = 1$), Eq. 2 simplifies to 
\begin{equation}
    \frac{s_\text{out}}{s_\text{in}} = -r-\frac{t^2\phi_0\phi_1}{1+r\phi_0\phi_1}
\end{equation}
In this special condition, a supermode emerges with an effective roundtrip propagation factor of $\phi_0\phi_1 = \alpha_{\text{rt},0}\alpha_{\text{rt},1} e^{\mathrm{i}\frac{\omega}{c}(n_{\text{eff},0}+n_{\text{eff},1})L}$. The corresponding free spectral range for this supermode is $\Delta\omega_\text{FSR} = \frac{2\pi c}{(n_{\text{g},0}+n_{\text{g},1})L}$. When both transverse modes' group indices are similar $n_{\text{g},0} \approx n_{\text{g},1}$, the FSR is approximately halved, aligning well with coupled-mode theory. For the generalized case involving $M$ transverse modes (illustrated for the three-mode scenario in Fig. \ref{fig:fig4}(a)), where each transverse mode is cascaded and fully converted into the next, the supermode has an effective roundtrip propagation factor 
\begin{equation}
    \Phi_\text{super} = \prod_{m=0}^{M-1} \phi_m
\end{equation}
The FSR of the resulting supermode is given by $\Delta\omega_\text{FSR} = \frac{2\pi c}{\sum_{m=0}^{M-1} n_{\text{g},m}L}$ (Sec. S4). 

We further conduct finite difference time domain simulations of the complete single-ring photonic molecule device to investigate the dependence of the resonance frequency splitting on the grating period number $N_1$ (Fig. \ref{fig:fig3}(b), Figs. S8 and S9, Sec. S5). The results reveal that the resonance frequencies initially split and subsequently converge, forming a distinct diamond-shaped pattern, as illustrated in Fig. \ref{fig:fig3}(b). This characteristic behavior occurs due to variations in the power conversion efficiency $\eta$, which increases with $N_1$ until reaching a maximum, and then decreases. Consequently, the optical power periodically transfers between the two transverse modes. Because the gratings on the ring waveguide are designed in an additive manner, increasing the grating period number $N_1$ effectively widens the ring waveguide and raises the modes' effective indices. This red-shifts the resonance frequencies and tilts the diamond-shaped pattern. To adjust this shift, gratings can be designed in a subtractive manner to blue-shift, or in additive-subtractive pairs to minimize the overall shift. Furthermore, the resonance frequency difference $\omega_\text{diff}$ between the $\mathrm{TE}_0$ and $\mathrm{TE}_1$ modes exhibits periodic variations as the input light frequency changes (Fig. S10).

Next, we experimentally demonstrate the splitting behavior of the real and imaginary eigenvalue components in the single-ring photonic molecule system. We manufactured experimental devices from 220-nm thick silicon-on-insulator material (Sec. S5). Fig. \ref{fig:fig3}(a) shows one representative device of a device series (Table S1) with varying grating period numbers ($N_1$) and constant corrugation depth of $h_1 = 20$ nm. Due to fabrication imperfections and material property variations across the chip, devices exhibit random resonance wavelength shifts. \revision{To enable consistent comparison of the splitting resonances across devices, we shift each measured spectrum along the wavelength axis so that its reference resonance, which is the higher frequency member of the split resonance pair closest to the target wavelength (1525 nm), is aligned to 1525 nm}, as shown in Fig. \ref{fig:fig3}(c) and Fig. S12. The unshifted spectra are provided in Fig. S11.

The measured normalized transmission spectra clearly illustrate the resonance frequency (wavelength) splitting from a single resonance to a dual resonance as $N_1$ varies. The splitting periodically increases and subsequently decreases as $N_1$ continues to grow, forming diamond-shaped patterns (Fig. \ref{fig:fig3}(c)). The deviation between the experimental and simulation results arises because the fabricated grating corrugation depth is larger than designed, due to fabrication resolution limitations. The resonance frequency splitting spans nearly the entire half-FSR range, from zero up to approximately $\frac{1}{2}\Delta\omega_\text{FSR}$ (Fig. \ref{fig:fig3}(d)). Achieving very small resonance splittings is challenging, as indicated by the sparse data around $2\text{Re}(\sigma)/\Delta\omega_{\text{FSR}} = 0$ \revision{(highlighted as purple area in Fig. \ref{fig:fig3}(d))}. To illustrate the effects of corrugation depth, additional devices were fabricated with depths $h_1 = 10$ nm (Fig. S13) and $h_1 = 35$ nm (Fig. S14). This challenge arises primarily because the power conversion efficiency ($\eta$) cannot be continuously adjusted and instead varies discretely with $N_1$. The minimally achievable increment is $\Delta\eta_\text{min} = \sin^2(\kappa\Lambda_1)$ (when assuming zero phase mismatch $\delta = 0$). Reducing the corrugation depth $h_1$ decreases the coupling coefficient $\kappa$ and reduces $\Delta\eta_\text{min}$. However, fabrication limitations place practical constraints on achieving very small values of $h_1$.

Moreover, frequency-dependent behaviors of both real (Fig. \ref{fig:fig3}(e), blue dots) and imaginary (Fig. \ref{fig:fig3}(e), red dots) components of eigenvalue splitting are experimentally observed. The real part corresponds directly to resonance frequency splitting, while the imaginary part manifests as splitting of the intrinsic quality factor (Fig. \ref{fig:fig3}(f), and further detailed in Figs. S16–S22). We also observe signatures of exceptional points (Fig. \ref{fig:fig3}(f) and Fig. S17). The experimental observations align closely with the predictions of coupled-mode theory, as indicated by the yellow fitting curve in Fig. \ref{fig:fig3}f, \revision{with the exception of the low-frequency region. At these frequencies, the device becomes deeply undercoupled (Fig. S11), leading to a vanishing resonance contrast in the transmission spectrum and an apparent saturation of the extracted quality factor.}

Our system can be easily scaled up to form a photonic trimer or larger cluster by cascading additional TMCs to generate and couple more transverse modes within the same single ring (Fig. \ref{fig:fig4}(a)). We design a TMC using symmetric gratings with a corrugation depth $h_2 = 15$ nm and grating period $\Lambda_2 = 2250$ nm, which selectively couples the $\mathrm{TE}_0$ and $\mathrm{TE}_2$ modes. The simulated power conversion efficiency $\eta_{02}$ of this TMC as a function of grating period number $N_2$ is shown in Fig. \ref{fig:fig4}(b). By adding this $\mathrm{TE}_0$-$\mathrm{TE}_2$ TMC into the single-ring photonic molecule system, we obtain a photonic trimer. Its simulated transmission and power spectra for different transverse modes are shown in Fig. \ref{fig:fig4}(c) and Fig. S23. We observe more complex resonance frequency splitting behavior in this photonic trimer. Specifically, the resonances can split from one into three, and the FSR can be approximately trisected ($\frac{1}{3}\Delta\omega_\text{FSR}$).

We demonstrated a multimode single-ring photonic molecule with controllable inter-mode coupling, a novel building block for integrated photonics. By harnessing multiple transverse modes within a single cavity and engineering their interactions via TMCs, we achieve precise control over the resonance splittings and intrinsic losses. This compact, scalable approach overcomes the geometric limitations of traditional multi-cavity systems (see a comparison in Table S2, Sec. S6). The demonstrated ability to generate bright–dark mode pairs, tune exceptional point dynamics, and independently control the real and imaginary parts of the system eigenvalues offers new capabilities in light–matter interaction control. This platform can be extended to implement complex multilevel energy structures \cite{hodaei2017enhanced}, synthetic frequency dimensions \cite{hu2020realization}, and non-Hermitian dynamics \cite{reisenbauer2024non}. Integration with active tuning elements such as thermo-optic or electro-optic modulators could further support dynamic reconfiguration. Overall, our approach lays the groundwork for a new generation of programmable, multimode photonic molecules for classical and quantum technologies.

We acknowledge support from AFOSR grants FA9550-23-1-0699. The fabrication of this work was performed at the Center for Nanoscale Systems (CNS) of Harvard University, which is supported by the NSF under award no. ECCS-2025158. I.-C.B.-C. acknowledges support from PRIMA Grant number 201547 from the Swiss National Science Foundation. V.G. acknowledges support from Research Foundation Flanders under grant numbers G032822N and G0K9322N. M.O. acknowledges funding by the European Union (grant agreement 101076933 EUVORAM). The views and opinions expressed are, however, those of the author(s) only and do not necessarily reflect those of the European Union or the European Research Council Executive Agency. Neither the European Union nor the granting authority can be held responsible for them.

\bibliography{srpm}
\end{document}